\documentclass[a4paper,fleqn,usenatbib]{mnras}

\usepackage[T1]{fontenc}
\usepackage{ae,aecompl}

\usepackage{graphicx}	
\usepackage{amsmath}	
\usepackage{amssymb}	
\usepackage{subfigure}
\usepackage{multirow}

\newcommand{\pcc}{\,{\rm cm}^{-3}}

\newcommand{\pcs}{\,{\rm cm}^{-2}}

\newcommand{\kel}{\, {\rm K}}

\newcommand{\msun}{\, {\rm M}_\odot}
\newcommand{\nh}{n_{\rm H}}

\newcommand{\pc}{\, {\rm pc}}

\newcommand{\yr}{\, {\rm yr}}
\newcommand{\myr}{\, {\rm Myr}}

\newcommand{\kms}{\, {\rm km \, s^{-1}}}

\title[Star formation line profiles]{Investigating the role of magnetic fields in star formation using molecular line profiles}

\author[Yin et al.]{Charles Yin$^{1}$\thanks{charlesyin42@gmail.com}, Felix D. Priestley$^{1}$ \& James Wurster$^{2}$
\\
$^{1}$School of Physics and Astronomy, Cardiff University, Queen's Buildings, The Parade, Cardiff CF24 3AA, UK \\
$^{2}$SUPA, School of Physics and Astronomy, University of St. Andrews, North Haugh, St Andrews, Fife KY16 9SS, UK \\
}

\date{Accepted XXX. Received YYY; in original form ZZZ}

\pubyear{2020}

\begin{document}
\label{firstpage}
\pagerange{\pageref{firstpage}--\pageref{lastpage}}
\maketitle

\begin{abstract}

Determining the importance of magnetic fields in star forming environments is hampered by the difficulty of accurately measuring both field strength and gas properties in molecular clouds. We post-process three-dimensional non-ideal magnetohydrodynamic simulations of prestellar cores with a time-dependent chemical network, and use radiative transfer modelling to calculate self-consistent molecular line profiles. Varying the initial mass-to-flux ratio from sub- to super-critical results in significant changes to both the intensity and shape of several observationally important molecular lines. We identify the peak intensity ratio of N$_2$H$^+$ to CS lines, and the CS $J=2-1$ blue-to-red peak intensity ratio, as promising diagnostics of the initial mass-to-flux ratio, with N$_2$H$^+$/CS values of $>0.6$ ($<0.2$) and CS blue/red values of $<3$ ($>5$) indicating subcritical (supercritial) collapse. These criteria suggest that, despite presently being magnetically supercritical, L1498 formed from subcritical initial conditions.

\end{abstract}




\section{Introduction}

The importance of magnetic fields in the collapse of prestellar cores to protostars is still poorly understood. Broadly speaking, theories of star formation can be separated into those that assume prestellar cores form above the critical mass for gravitational collapse \citep{mouschovias1976}, and those that predict that cores are initially subcritical, and contract until the reduction in magnetic flux due to ambipolar diffusion allows the central region to collapse \citep{fiedler1993}. In the former case, star formation occurs on the free-fall timescale, $\sim 10^6 \yr$ under typical molecular cloud conditions, whereas in the latter the relevant timescale is that of ambipolar diffusion, around an order of magnitude larger \citep{tassis2004a,banerji2009}. While there are other effects which have an impact on the collapse timescale, such as the cosmic ray ionisation rate, these effects are typically less significant than that of the initial mass-to-flux ratio \citep{wurster2018a}.

Collapse timescales not only affect the rate of conversion of interstellar material into stars, but also change the chemical composition of the infalling gas and ice mantles \citep{tassis2012,priestley2018}, which is inherited, at least in part, by the subsequent protoplanetary disc \citep{oberg2020,coutens2020}. The mechanism of prestellar collapse can thus have consequences on both smaller and larger scales than of the cores themselves.

Direct measurements of the magnetic field strength in molecular clouds typically find values which favour supercritical models of star formation \citep{crutcher2009,crutcher2012}, although some exceptions exist \citep[e.g.][]{soam2019}. However, values for the mass-to-flux ratio for prestellar objects are often close to the critical value and have large uncertainties \citep{soam2018,beltran2019}. Concerns have also been raised about the reliability of magnetic field strengths measured via the Zeeman effect and their relation to the gas density, due to assumptions about the abundance of the OH molecule \citep{tassis2014} and the statistical methods employed \citep{tritsis2015,jiang2020}.

This has led to alternative tests being proposed, typically exploiting the effect of the increased timescale in initially subcritical models on the molecular abundances, which can be determined relatively easily from observations. \citet{lippok2013} used the CO depletion in a sample of starless cores to measure chemical ages $\lesssim 1 \myr$, suggesting supercritical collapse, while \citet{pagani2013} reached a similar conclusion for L183 based on deuteration measurements of N$_2$H$^+$. \citet{lin2020} found that deuteration profiles imply an age $> 1 \myr$ for L1512, which is more typical of ambipolar diffusion models. \citet{tassis2012} used coupled hydrodynamical-chemical models to identify various molecular abundance ratios which are sensitive to the collapse timescale, although these models are based on the one-dimensional thin disc approximation.

In \citet{priestley2019}, we post-processed a fully three-dimensional non-ideal magnetohydrodynamical (MHD) model with a time-dependent chemical network in order to determine the molecular structure of initially sub- and supercritical collapse models. While several molecules differ by orders of magnitude in abundance in the central regions of the prestellar cores, due to enhanced freeze-out in the subcritical models, we found the molecular column density profiles were too similar to distinguish the two cases, due to intervening material along the line of sight. In this paper, we instead focus on the synthetic line profiles of several commonly observed species, taking advantage of the velocity structure to probe regions where subcritical and supercritical models of collapse are clearly distinct.

 \section{Method}

\begin{table*}
  \centering
  \caption{Initial radius, mass density, hydrogen nuclei density, magnetic field strength, duration and mass-to-flux ratios for the LOW-SUP and LOW-SUB non-ideal MHD models. $\lambda$ is the mass to flux ratio.}
  \begin{tabular}{cccccccc}
    \hline
    Model   & $R$ / $\pc$ & $\rho$ / g $\pcc$      & $\nh$ / $\pcc$     & $B_z$ / $\mu$G & $t_{\rm end}$ / Myr & $\lambda$ / $\lambda_{crit}$ \\ \hline
    LOW-SUP & $0.13$      & $3.68 \times 10^{-20}$ & $1.57 \times 10^4$ & $9.1$          & $0.347$             & 5                            \\
    LOW-SUB & $0.13$      & $3.68 \times 10^{-20}$ & $1.57 \times 10^4$ & $91$           & $0.972$             & 0.5                          \\ \hline
  \end{tabular}
  \label{tab:cloudprop}
\end{table*}

\begin{figure*}
	\subfigure{\includegraphics[width=0.45\textwidth]{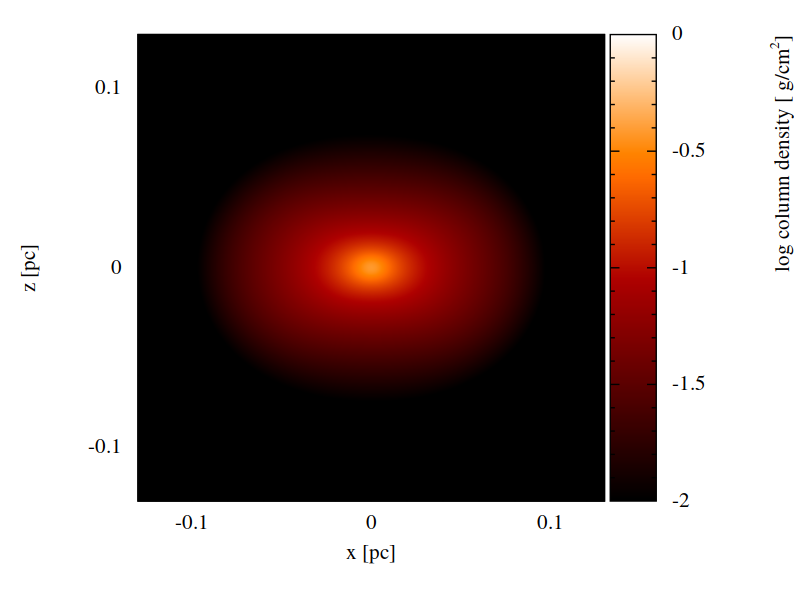}}\quad
	\subfigure{\includegraphics[width=0.45\textwidth]{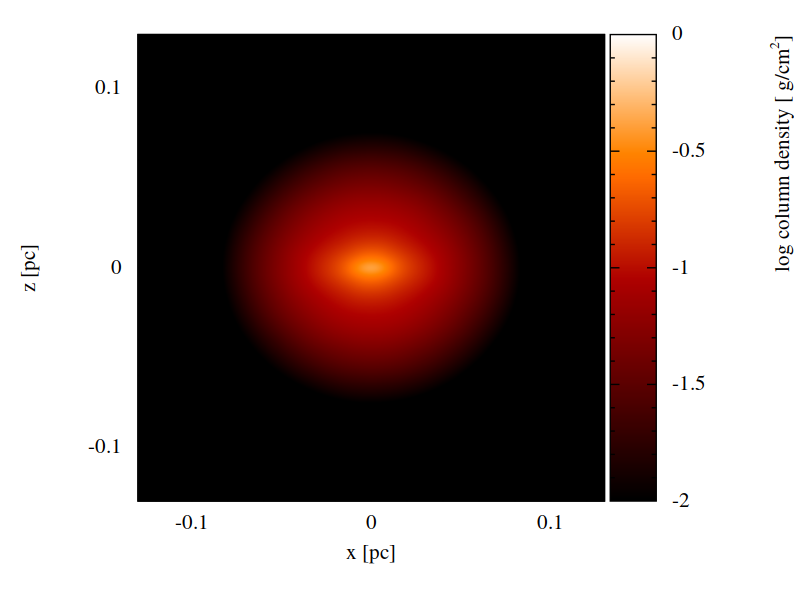}}
	\caption{Column density of the LOW-SUB (left) and LOW-SUP (right) models, viewed side-on.}
	\label{fig:dens}
\end{figure*}

Following \citet{priestley2019}, we use the {\sc phantom} smoothed particle (magneto)hydrodynamics (SPH) code \citep{price2018} to run models of spherical, static, uniform density prestellar cores with a constant magnetic field in the $z$-direction. The ambipolar diffusion coefficient is calculated using the NICIL library \citep{wurster2016}; as NICIL does not include molecular ions, which are the dominant ionised species at the densities we investigate, we assume the ion density is given by
 \begin{equation}
    n_i/\nh = 10^{-7} \left( \frac{\nh}{10^3 \pcc} \right)^{-0.6},
 \end{equation}
  which accurately reproduces the ion-neutral relation produced by full, time-dependent chemical networks to within a factor of a few over the relevant range of densities \citep{tassis2012b,priestley2019}.

We assume an isothermal equation of state with sound speed $c_s = 0.2 \kms$, corresponding to molecular gas at $\sim 10 \kel$. The cores are surrounded by a background medium with the density reduced by a factor of $30$, with the temperature increased by an equivalent factor to ensure pressure balance\footnote{The models in \citet{priestley2019} were erroneously run with a background medium temperature of $10 \kel$, affecting the core dynamics and subsequent chemical evolution. The authors of that paper have submitted an erratum; the results presented here have the correct background temperature. As the high-density models in \citet{priestley2019} resulted in anomalously low CO abundances compared to observed prestellar cores, we focus on the low-density case, with a core mass $M = 5 \msun$ and radius $R = 0.13 \pc$, giving an initial hydrogen nuclei density $\nh = 1.57 \times 10^4 \pcc$. We consider models with mass-to-flux ratios of $5$ (LOW-SUP) and $0.5$ (LOW-SUB) times the critical value \citep{mouschovias1976}, equivalent to initial magnetic field strengths $B_z = 9.1$ and $91 \, {\rm \mu G}$\footnote{These are slightly higher than in \citet{priestley2019} due to a minor change to the definition of the critical mass-to-flux ratio.}. The models are terminated when the central density reaches $\nh =2 \times 10^6 \pcc$, which occurs after $0.305$ and $0.964 \myr$ for the LOW-SUP and LOW-SUB models respectively. The models are run with $\sim 200 \, 000$ particles, giving a particle mass of $3.5 \times 10^{-5} \msun$. Model properties are listed in Table \ref{tab:cloudprop}. The final column density structures for both models, viewed side on, are shown in Figure \ref{fig:dens}.}

We post-process a randomly-selected subset of $10 \, 000$ particles with the {\sc UCLCHEM} chemical evolution code \citep{holdship2017} to obtain the three-dimensional molecular structure of each model core, using the UMIST12 reaction rate network \citep{mcelroy2013} and the low-metal elemental abundances from \citet{lee1998}. We assume a constant gas/grain temperature $10 \kel$ and cosmic ray ionisation rate $1.3 \times 10^{-17} \, {\rm s^{-1}}$, and set the radiation field to zero; prestellar cores are typically located within larger molecular clouds, and as such are shielded from any external radiation field.

We use LIME \citep{brinch2010} to calculate line intensities from our MHD-chemical modelling results for a number of observationally important molecules, taking molecular data from the LAMDA database \citep{schoier2005}. Density, molecular abundance and other properties are assigned to sampling points from the nearest-neighbour SPH particle which was post-processed chemically. We use $10 \, 000$ spatial grid points and {\bf 201} velocity channels with a spacing of $0.01 \kms$, which we have confirmed are sufficient that our results are converged. We assume isotopic ratios of ${\rm ^{12}C/^{13}C}=100$ and ${\rm ^{16}O/^{18}O}=500$. We evaluate line profiles at the simulation end-points (central density $\nh = 2 \times 10^6 \pcc$) by averaging the line intensity over all pixels within a radius $0.13 \pc$ of the centre.

\section{Results}

\begin{figure*}
	\includegraphics[width=\textwidth]{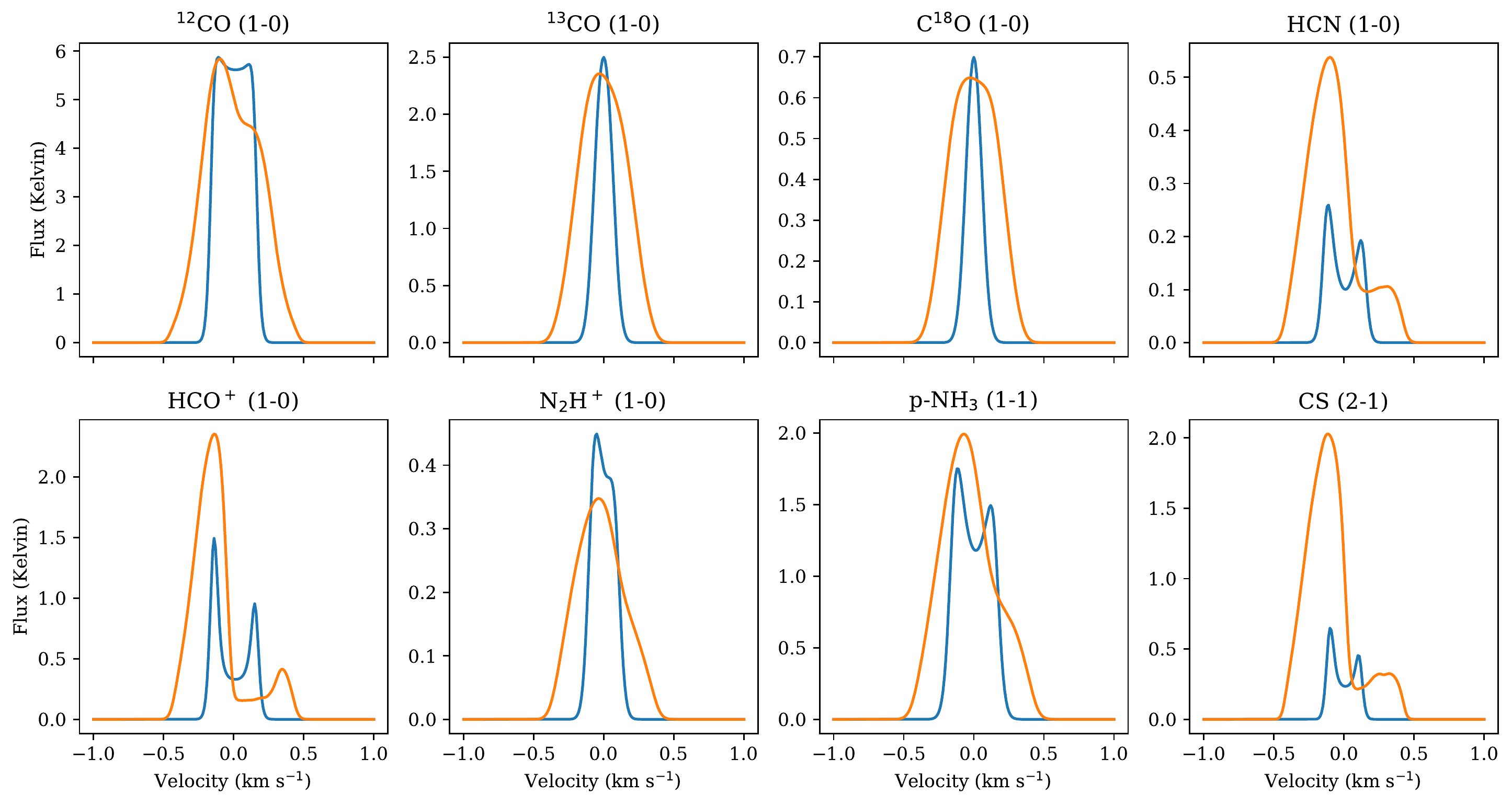}
	\caption{Line profiles for various species viewed side-on. Each species is labelled above the plot. The blue line is the LOW-SUB model, while the orange line is the LOW-SUP model.}	\label{fig:xzlines}
\end{figure*}

\begin{figure*}
	\includegraphics[width=\textwidth]{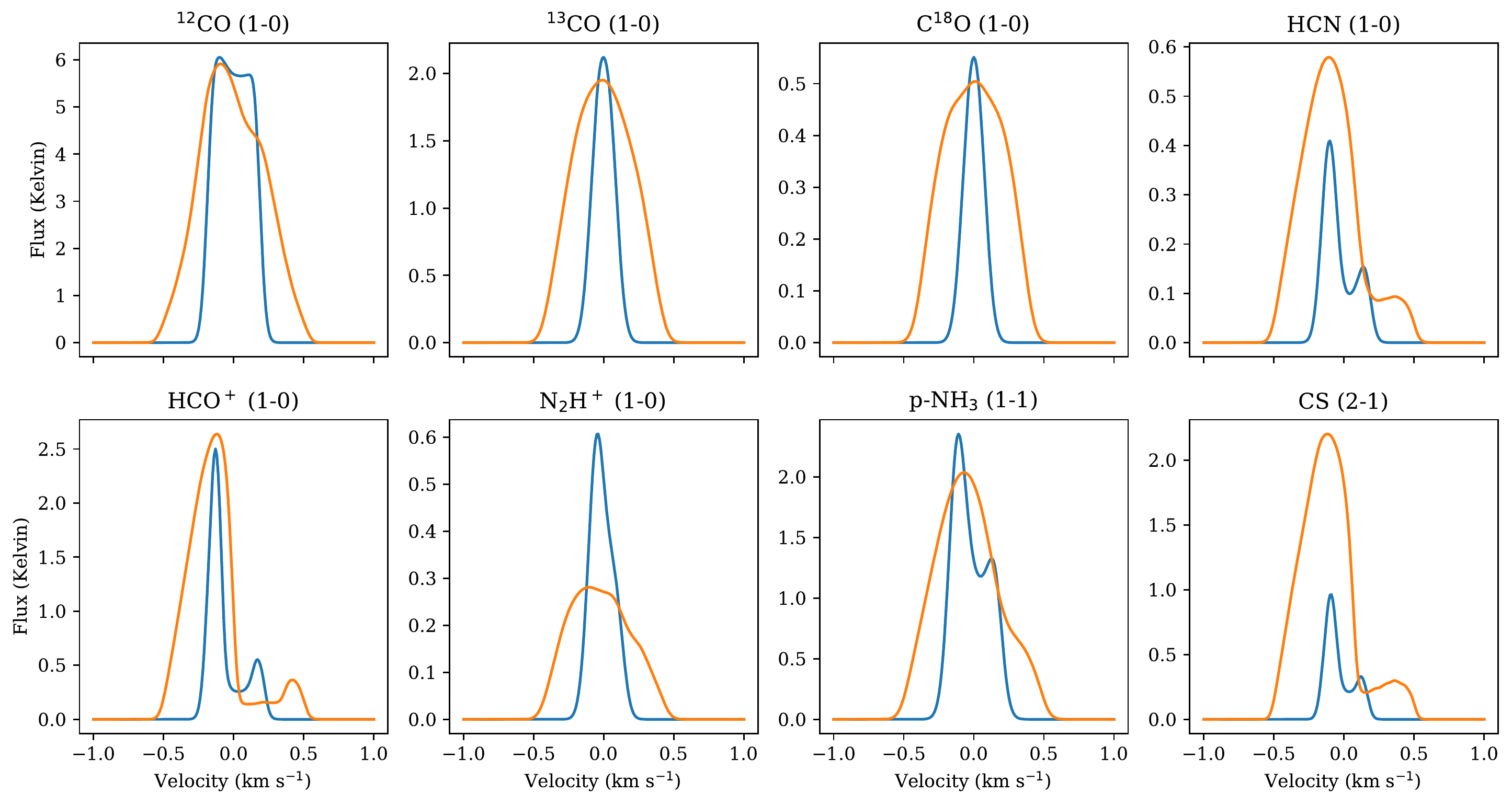}
	\caption{Line profiles for various species viewed face-on. Each species is labelled above the plot. The blue line is the LOW-SUB model, while the orange line is the LOW-SUP model.}
	\label{fig:xylines}
\end{figure*}

\begin{figure*}
	\includegraphics[width=\textwidth]{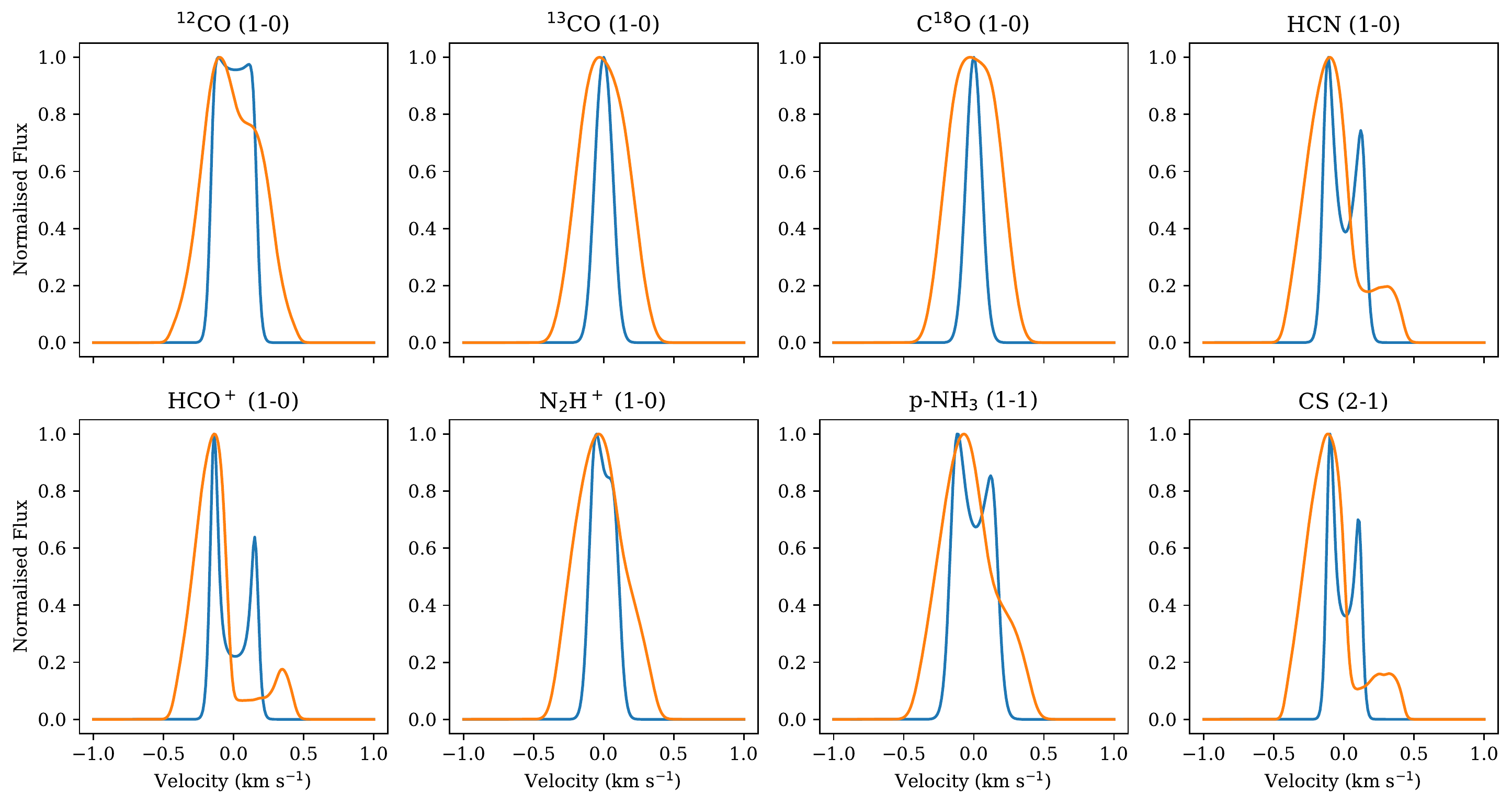}
	\caption{Normalized line profiles for various species viewed side-on. Each species is labelled above the plot. The blue line is the LOW-SUB model, while the orange line is the LOW-SUP model.}
	\label{fig:xzlinesnorm}
\end{figure*}

\begin{figure*}
	\includegraphics[width=\textwidth]{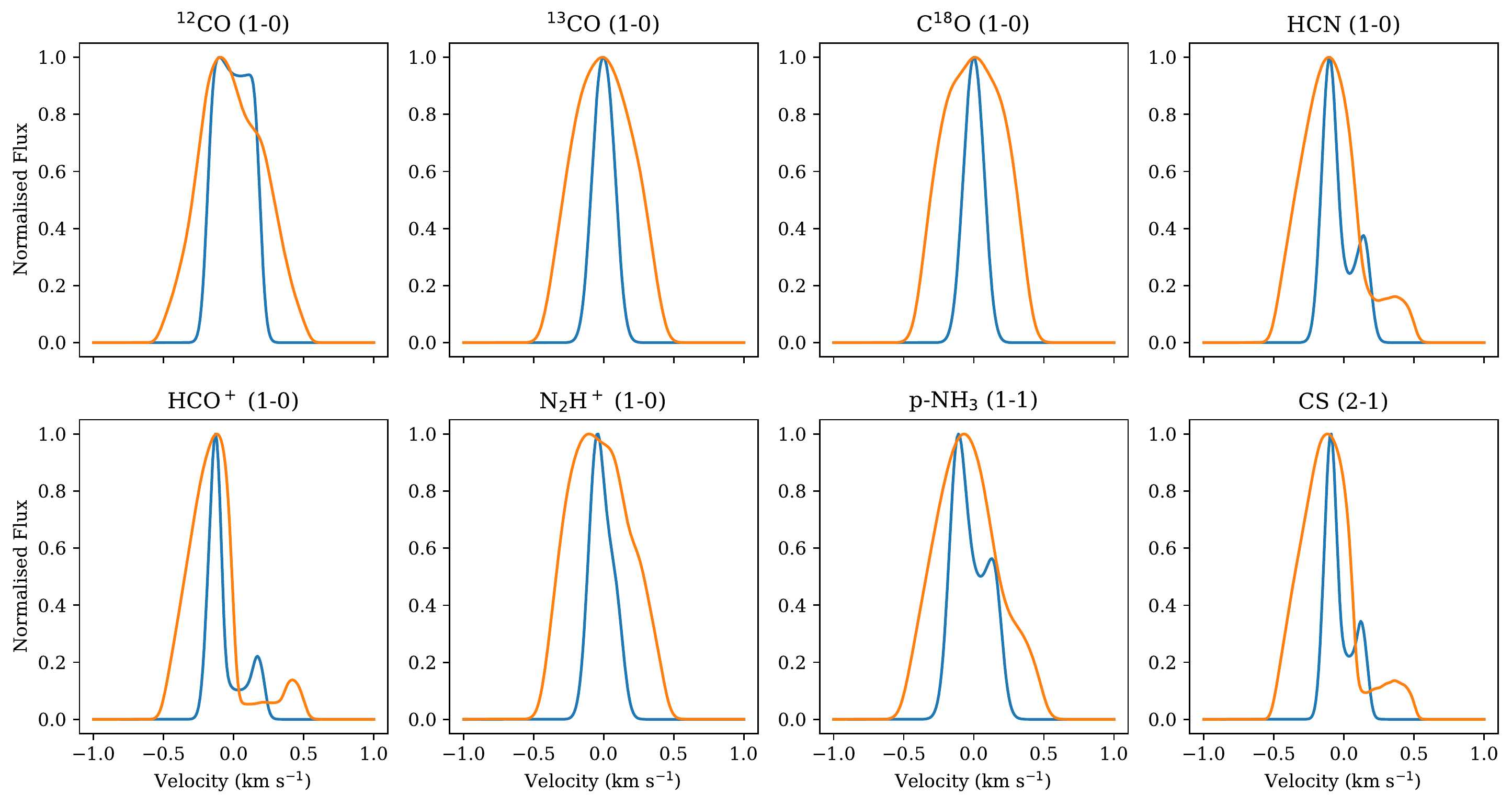}
	\caption{Normalized line profiles for various species viewed face-on. Each species is labelled above the plot. The blue line is the LOW-SUB model, while the orange line is the LOW-SUP model.}
	\label{fig:xylinesnorm}
\end{figure*}

\begin{table}
	\centering
	\caption{Parameters for one or two Gaussian curves fitted to the line profile of various species when viewed side-on. For each species, first row is the LOW-SUB model, while the second row is the LOW-SUP model.}
	\begin{tabular}{@{}lllllll@{}}
	\hline
	              & \multicolumn{3}{|c|}{Gaussian 1}           & \multicolumn{3}{|c|}{Gaussian 2}         \\ \hline
	              & Height & Centre          & Width           & Height & Centre          & Width           \\
	              & $\kel$ & $\rm m\,s^{-1}$ & $\rm m\,s^{-1}$ & $\kel$ & $\rm m\,s^{-1}$ & $\rm m\,s^{-1}$ \\ \hline
	$\rm ^{12}CO$ & 6.10   & -87.5           & 63.5            & 6.07   & 83.4            & 66.3            \\
                  & 5.67   & -108            & 134             & 3.59   & 181             & 112             \\ \hline
	$\rm ^{13}CO$ & 2.55   & -0.270          & 61.3            & -      & -               & -               \\
	              & 2.48   & -4.13           & 170             & -      & -               & -               \\ \hline
	$\rm C^{18}O$ & 0.702  & -0.0533         & 55.6            & -      & -               & -               \\
	              & 0.707  & \ 0.213         & 173             & -      & -               & -               \\ \hline
	HCN           & 0.248  & -102            & 489             & 0.179  & 97.8            & 59.1            \\
	              & 0.547  & -125            & 138             & 0.111  & 303             & 81.7            \\ \hline
	$\rm HCO^+$   & 1.38   & -134            & 36.2            & 0.659  & 115             & 74.0            \\
	              & 2.38   & -178            & 104             & 0.369  & 323             & 88.4            \\ \hline
	$\rm N_2H^+$  & 0.430  & -57.5           & 466             & 0.362  & 57.5            & 49.1            \\
	              & 0.355  & -48.5           & 161             & 0.0587 & 251             & 78.5            \\ \hline
	p-NH$_3$      & 1.74   & -101            & 63.7            & 1.50   & 97.6            & 69.7            \\
	              & 2.00   & -80.7           & 164             & 0.493  & 284             & 90.4            \\ \hline
	CS            & 0.609  & -89.3           & 361             & 0.387  & 78.0            & 54.3            \\ 
	              & 2.07   & -141            & 123             & 0.365  & 297             & 83.2            \\ \hline
	\end{tabular}
	\label{tab:xzfitparam}
\end{table}

\begin{table}
	\centering
	\caption{Parameters for one or two Gaussian curves fitted to the line profile of various species when viewed from the $z$ axis. For each species, first row is the LOW-SUB model, while the second row is the LOW-SUP model.}
	\begin{tabular}{lllllll}
	\hline
	              & \multicolumn{3}{|c|}{Gaussian 1}           & \multicolumn{3}{|c|}{Gaussian 2}           \\ \hline
	              & Height & Centre          & Width           & Height & Centre          & Width           \\
	              & $\kel$ & $\rm m\,s^{-1}$ & $\rm m\,s^{-1}$ & $\kel$ & $\rm m\,s^{-1}$ & $\rm m\,s^{-1}$ \\ \hline
	$\rm ^{12}CO$ & 6.23   & -95.1           & 75.3            & 5.87   & 97.5            & 74.3            \\
                  & 5.74   & -109            & 168             & 3.02   & 223             & 136             \\ \hline
	$\rm ^{13}CO$ & 2.18   & -1.33           & 76.4            & -      & -               & -               \\
	              & 2.05   & -4.79           & 222             & -      & -               & -               \\ \hline
	$\rm C^{18}O$ & 0.558  & -0.450          & 71.6            & -      & -               & -               \\
	              & 0.544  & \ 1.27          & 231             & -      & -               & -               \\ \hline
	HCN           & 0.407  & -101            & 57.3            & 0.151  & 120             & 65.1            \\
	              & 0.594  & -127            & 178             & 0.0903 & 391             & 780             \\ \hline
	$\rm HCO^+$   & 2.47   & -133            & 46.0            & 0.474  & 139             & 79.5            \\
	              & 2.68   & -176            & 137             & 0.340  & 404             & 84.1            \\ \hline
	$\rm N_2H^+$  & 0.589  & -49.4           & 60.0            & 0.256  & 79.3            & 54.9            \\
	              & 0.244  & 60.5            & 197             & 0.167  & -218            & 122             \\ \hline
	p-NH$_3$      & 2.33   & -105            & 68.8            & 1.35   & 110             & 77.5            \\
	              & 2.07   & -76.1           & 211             & 0.353  & 371             & 86.6            \\ \hline
	CS            & 0.953  & -93.3           & 47.2            & 0.316  & 103             & 60.4            \\ 
	              & 2.28   & -144            & 161             & 0.314  & 380             & 80.9            \\ \hline
	\end{tabular}
	\label{tab:xyfitparam}
\end{table}

\begin{table}
	\centering
	\caption{Ratio of the intensity of blue to red peaks. The first line lists the ratio for the LOW-SUB model, and the second line lists the ratio for LOW-SUP model.}
	\begin{tabular}{lcc}
		Species       & Side On & $z$ Direction \\ \hline
		$\rm ^{12}CO$ & 1.00    & 1.06          \\
		              & 1.58    & 1.90          \\ \hline
		$\rm ^{13}CO$ & -       & -             \\
		              & -       & -             \\ \hline
		$\rm C^{18}O$ & -       & -             \\
		              & -       & -             \\ \hline
		HCN           & 1.38    & 2.70          \\
		              & 4.93    & 6.58          \\ \hline
		HCO$^+$       & 2.09    & 5.21          \\
		              & 6.45    & 7.88          \\ \hline
		N$_2$H$^+$    & 1.19    & 2.30          \\
		              & 6.05    & 0.684         \\ \hline
		p-NH$_3$      & 1.16    & 1.73          \\
		              & 4.06    & 5.86          \\ \hline
		CS            & 1.57    & 3.02          \\
		              & 5.67    & 7.26          \\ \hline
	\end{tabular}
	\label{tab:peakratio}
\end{table}

\begin{figure*}
  \centering
  \subfigure{\includegraphics[width=\columnwidth]{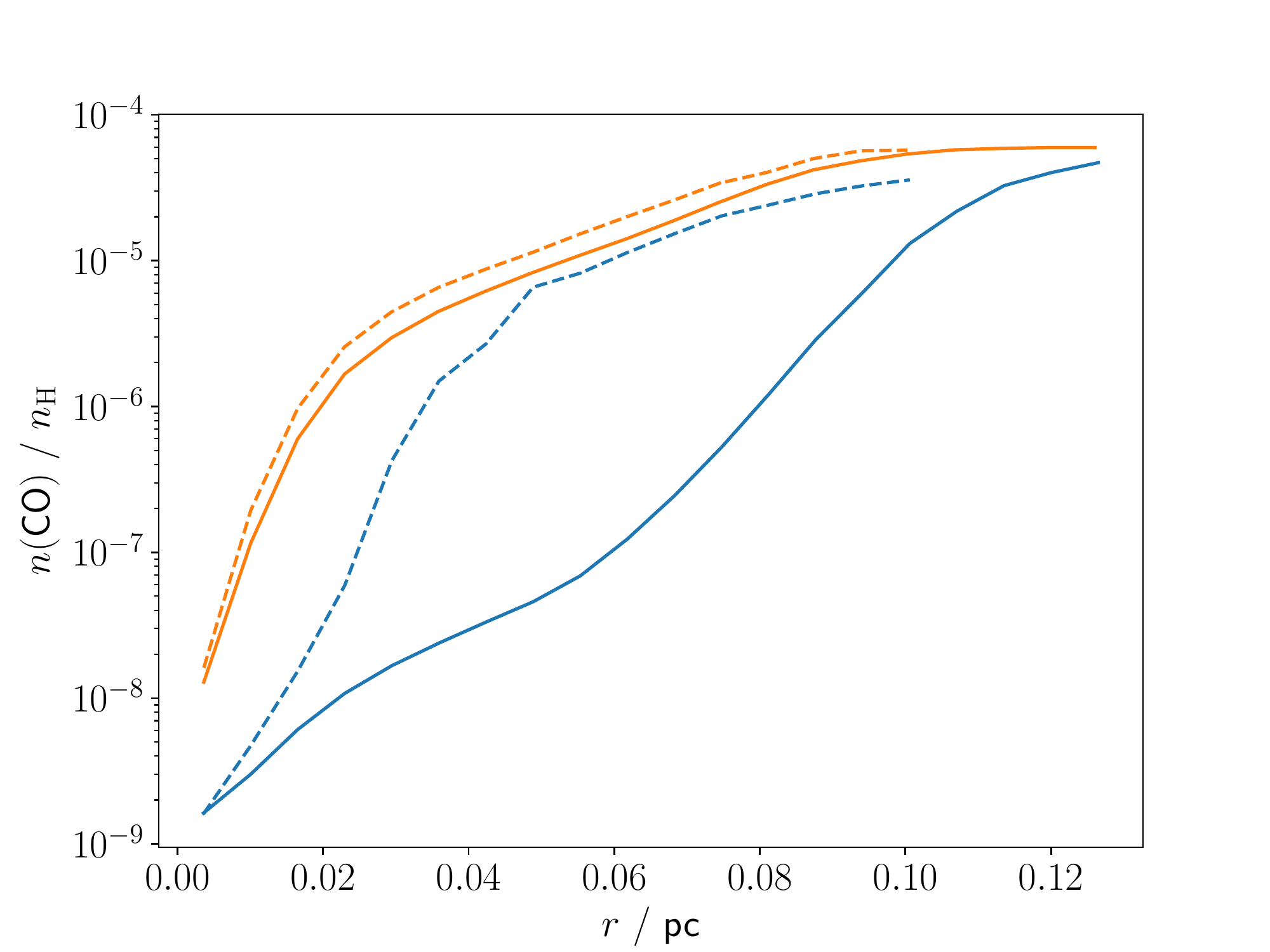}}\quad
  \subfigure{\includegraphics[width=\columnwidth]{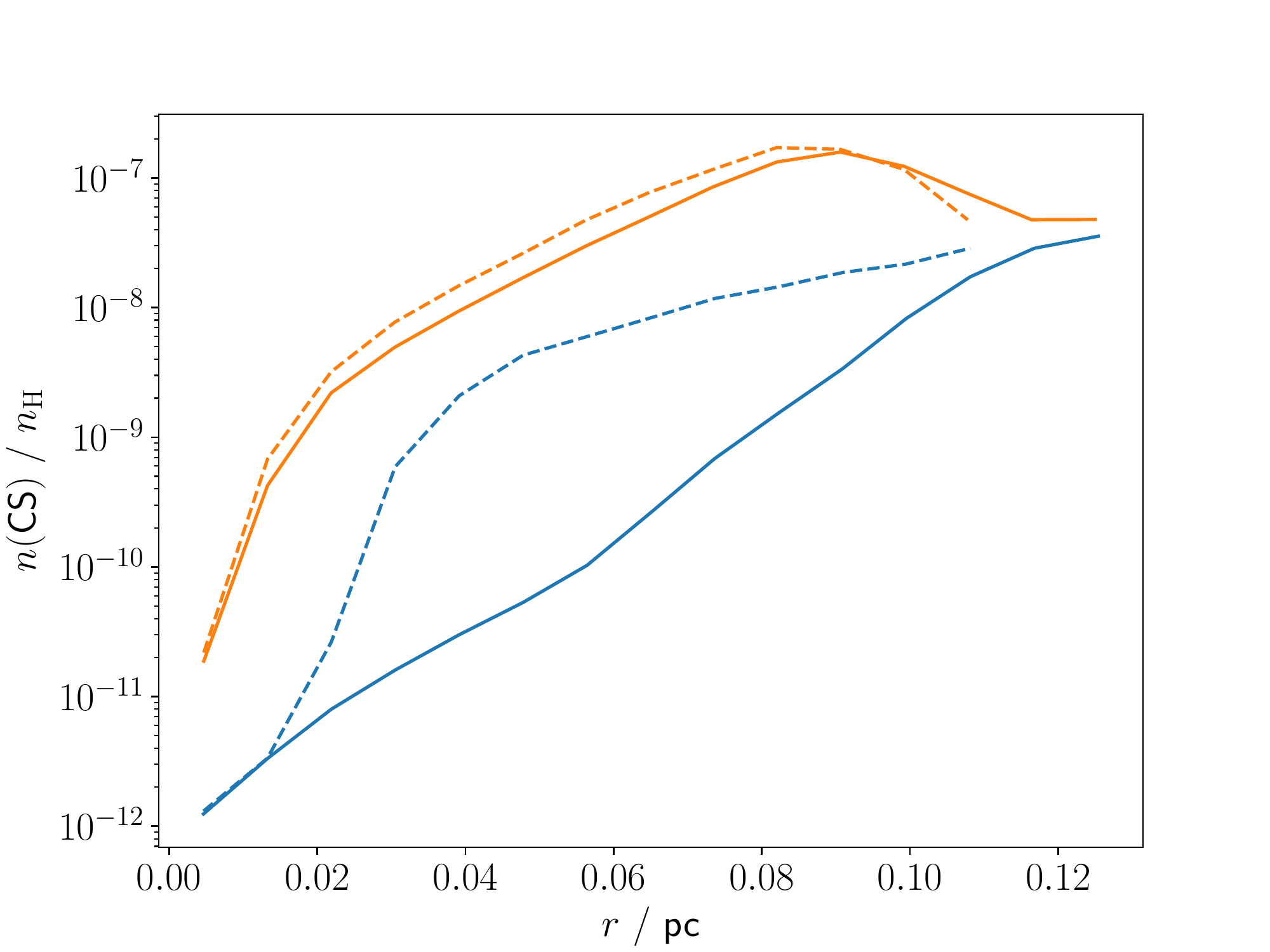}}\quad
  \caption{Midplane (solid lines) and $z$-axis (dashed lines) CO (left panel) and CS (right panel) abundance profiles for the LOW-SUP (orange) and LOW-SUB (blue) models.}
  \label{fig:abun}
\end{figure*}

\begin{table}
	\centering
	\caption{Peak intensity ratios of different lines when viewed side-on. For each ratio, the first line is the LOW-SUB model, and the second line is the LOW-SUP model. Each figure is the peak intensity of the species below divided by the species to its left.}
	\begin{tabular}{l@{\vline}ccccccc}
$^{13}$CO  & 2.35      &           &           &       &           &            &          \\
           & 2.48      &           &           &       &           &            &          \\\hline
C$^{18}$O  & 8.40      & 3.58      &           &       &           &            &          \\
           & 9.00      & 3.63      &           &       &           &            &          \\\hline
HCN        & 22.7      & 9.64      & 2.70      &       &           &            &          \\
           & 10.9      & 4.38      & 1.21      &       &           &            &          \\\hline
HCO$^{+}$  & 3.93      & 1.67      & 0.468     & 0.173 &           &            &          \\
           & 2.48      & 1.00      & 0.275     & 0.228 &           &            &          \\\hline
N$_2$H$^+$ & 13.1      & 5.56      & 1.56      & 0.577 & 3.33      &            &          \\
           & 16.8      & 6.78      & 1.87      & 1.55  & 6.78      &            &          \\\hline
p-NH$_3$   & 3.36      & 1.43      & 0.399     & 0.148 & 0.854     & 0.257      &          \\
           & 2.93      & 1.18      & 0.325     & 0.270 & 1.18      & 0.174      &          \\\hline
CS         & 9.06      & 3.86      & 1.078     & 0.400 & 2.31      & 0.693      & 2.70     \\
           & 2.88      & 1.16      & 0.320     & 0.265 & 1.16      & 0.171      & 0.983    \\\hline
	       & $^{12}$CO & $^{13}$CO & C$^{18}$O & HCN   & HCO$^{+}$ & N$_2$H$^+$ & p-NH$_3$ \\
	\end{tabular}
	\label{tab:abssubxz}
\end{table}

\begin{table}
	\centering
	\caption{Peak intensity ratios of different lines when viewed face-on. For each ratio, the first line is the LOW-SUB model, and the second line is the LOW-SUP model. Each figure is the peak intensity of the species below divided by the species to its left.}
	\begin{tabular}{l@{\vline}ccccccc}
$^{13}$CO  & 2.86      &           &           &       &           &            &          \\
           & 3.03      &           &           &       &           &            &          \\\hline
C$^{18}$O  & 11.0      & 3.85      &           &       &           &            &          \\
           & 11.7      & 3.87      &           &       &           &            &          \\\hline 
HCN        & 14.8      & 5.18      & 1.35      &       &           &            &          \\
           & 10.2      & 3.37      & 0.871     &       &           &            &          \\\hline
HCO$^{+}$  & 2.42      & 0.848     & 0.220     & 0.164 &           &            &          \\
           & 2.24      & 0.740     & 0.191     & 0.219 &           &            &          \\\hline
N$_2$H$^+$ & 9.98      & 3.49      & 0.908     & 0.674 & 4.12      &            &          \\
           & 21.0      & 6.94      & 1.79      & 2.059 & 9.39      &            &          \\\hline
p-NH$_3$   & 2.57      & 0.901     & 0.234     & 0.174 & 1.06      & 0.258      &          \\
           & 2.90      & 0.958     & 0.248     & 0.284 & 1.30      & 0.138      &          \\\hline
CS         & 6.28      & 2.20      & 0.571     & 0.425 & 2.59      & 0.629      & 2.44     \\
           & 2.69      & 0.886     & 0.229     & 0.263 & 1.20      & 0.128      & 0.925    \\\hline
	       & $^{12}$CO & $^{13}$CO & C$^{18}$O & HCN   & HCO$^{+}$ & N$_2$H$^+$ & p-NH$_3$ \\
	\end{tabular}
	\label{tab:abssubxy}
\end{table}

\begin{figure}
  \centering
  \includegraphics[width=\columnwidth]{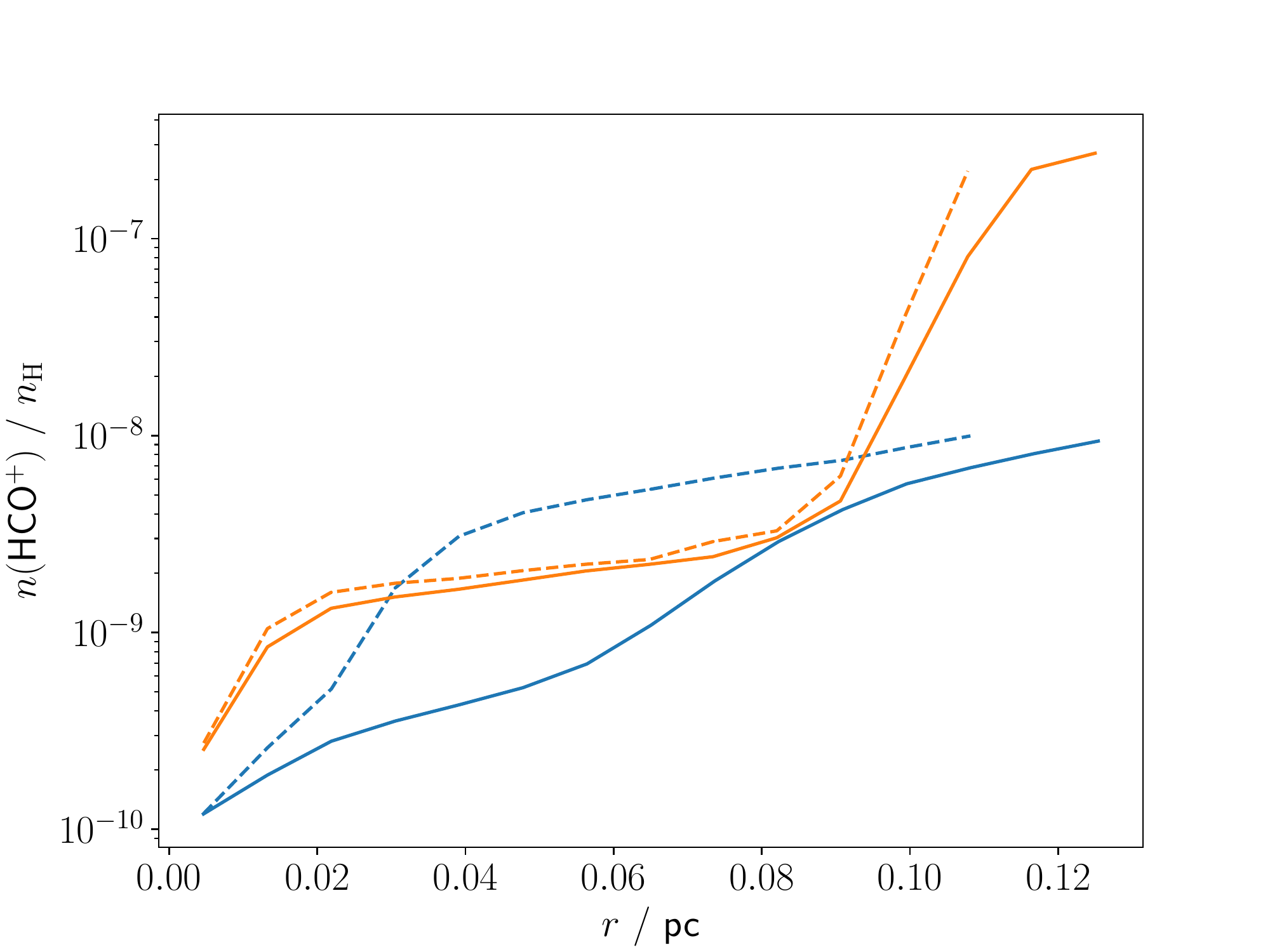}
  \caption{Midplane (solid lines) and $z$-axis (dashed lines) HCO$^+$ abundance profiles for the LOW-SUP (orange) and LOW-SUB (blue) models.}
  \label{fig:abun2}
\end{figure}

Figures \ref{fig:xzlines} and \ref{fig:xylines} show the average line profiles for a number of observationally-relevant molecules, for side-on ($x-z$ plane) and face-on ($x-y$ plane) orientations with respect to the magnetic field axis respectively. To better compare the shapes rather than the absolute intensities, Figures \ref{fig:xzlinesnorm} and \ref{fig:xylinesnorm} show the line profiles normalised to the peak intensity. Depending on whether the line appears singly- or doubly peaked, the profiles were fitted with either a single or double Gaussian profile, the fit parameters of which are given in Tables \ref{tab:xzfitparam} and \ref{tab:xyfitparam}. While the profiles are not necessarily Gaussian, this provides a simple way of comparing displacements and line widths.

All three CO isotopologues have similar peak intensities for the sub- and supercritical models, but the LOW-SUP line profiles are much broader, due to less depletion in the central, high-velocity regions, as can be seen in Figure \ref{fig:abun}. HCN and CS are both significantly weaker for LOW-SUB due to lower abundances throughout the core (CS is shown in Figure \ref{fig:abun}), whereas the p-NH$_3$ and N$_2$H$^+$ lines are of comparable or even greater strength in the LOW-SUB model as they are far less affected by depletion. HCO$^+$ shows the greatest variation with viewing angle; the midplane abundance (shown in Figure \ref{fig:abun2}) is lower, but the $z$-axis abundance higher, for the LOW-SUB model, resulting in the contrasting behaviour.

The asymmetric double-peaked appearance of some lines, with the blue peak stronger than the red, is due to self-absorption by inflowing material, i.e. an inverse P Cygni effect. This is commonly seen in observations of actual prestellar cores \citep{tafalla2002,tafalla2006}. For comparison, the line-centre optical depth for the LOW-SUB model viewed face-on is $\sim 0.5$ for $^{13}$CO (single peak) and $\sim 15$ for HCN (double-peaked). In both orientations, the LOW-SUP model results in broader lines. These effects are due to the larger abundances throughout the core in the faster-evolving supercritical model, where molecules have less time to freeze-out onto grain surfaces, as can be seen in Figure \ref{fig:abun} for CO and CS. However, these effects are unlikely to make useful diagnostics for the mechanism of collapse. The differences involved are not large (typically a factor of a few), and are likely to be strongly affected by the initial density and cloud size, along with other effects we do not consider such as rotation. We thus focus on differences in the properties of lines relative to each other, which should be much less sensitive to the initial conditions.

Table \ref{tab:peakratio} lists the ratios of blue to red peaks, as measured by our Gaussian fits. There is a significant different between the LOW-SUB and LOW-SUP models in the \textbf{$\rm ^{12}$CO, HCN, HCO$^+$, N$_2$H$^+$, p-NH$_3$ and CS} lines when viewed side on. However, when viewed face-on the difference is not present for \textbf{HCO$^+$}, and is reversed in \textbf{N$_2$H$^+$}. For these molecules, the abundance profiles face-on axis for the LOW-SUB model reach levels comparable to or larger than those in the LOW-SUP model at small radii, resulting in larger column densities and thus stronger absorption when viewed face-on. This can be seen for HCO$^+$ in Figure \ref{fig:abun2}. As cores may be observed at any inclination, peak ratios for these molecules cannot be used to discriminate between models. p-NH$_3$ shows blue-to-red peak intensity ratios of \textbf{$\sim 1.5$} for the LOW-SUP model and $\sim 4-6$ for the LOW-SUB model, but the red `peak' for the LOW-SUP model is barely detectable as such, appearing as more of a shoulder on the line profile. CS and HCN both have blue/red peak intensity ratios $\lesssim 3$ for the LOW-SUB model, and $\gtrsim 5$ for LOW-SUP, regardless of inclination, and are thus more promising as tracers of magnetic criticality.

In addition to considering the shape of the line, the ratios of peak intensities of the lines produced by different species are also potentially useful. These are given in Tables \ref{tab:abssubxz} and \ref{tab:abssubxy} for the side-on and face-on cases, respectively. Some differences are consistent between the two models regardless of orientation. These include \textbf{N$_2$H$^+$ and HCN; N$_2$H$^+$ and $\rm HCO^+$; $\rm ^{12}$CO and CS; $^{13}$CO and CS; C$^{18}$O and CS; and N$_2$H$^+$ and CS; N$_2$H$^+$ and CS.} These are all parings of species with a {\bf prominent} self-absorption feature with ones without. In particular, we note that the N$_2$H$^+$/CS and HCN/N$_2$H$^+$ peak intensity ratios are not greatly affected by viewing angle, but do vary significantly between the LOW-SUB and LOW-SUP models. The LOW-SUB model has N$_2$H$^+$/CS $>0.6$ and HCN/N$_2$H$^+$ $<1$, whereas the LOW-SUP model has N$_2$H$^+$/CS $<0.2$ and HCN/N$_2$H$^+$ $>1$, due to comparable N$_2$H$^+$ line strengths but significantly weaker CS/HCN intensity for the LOW-SUB model.

\section{Discussion}

\begin{figure*}
	\subfigure{\includegraphics[width=0.45\textwidth]{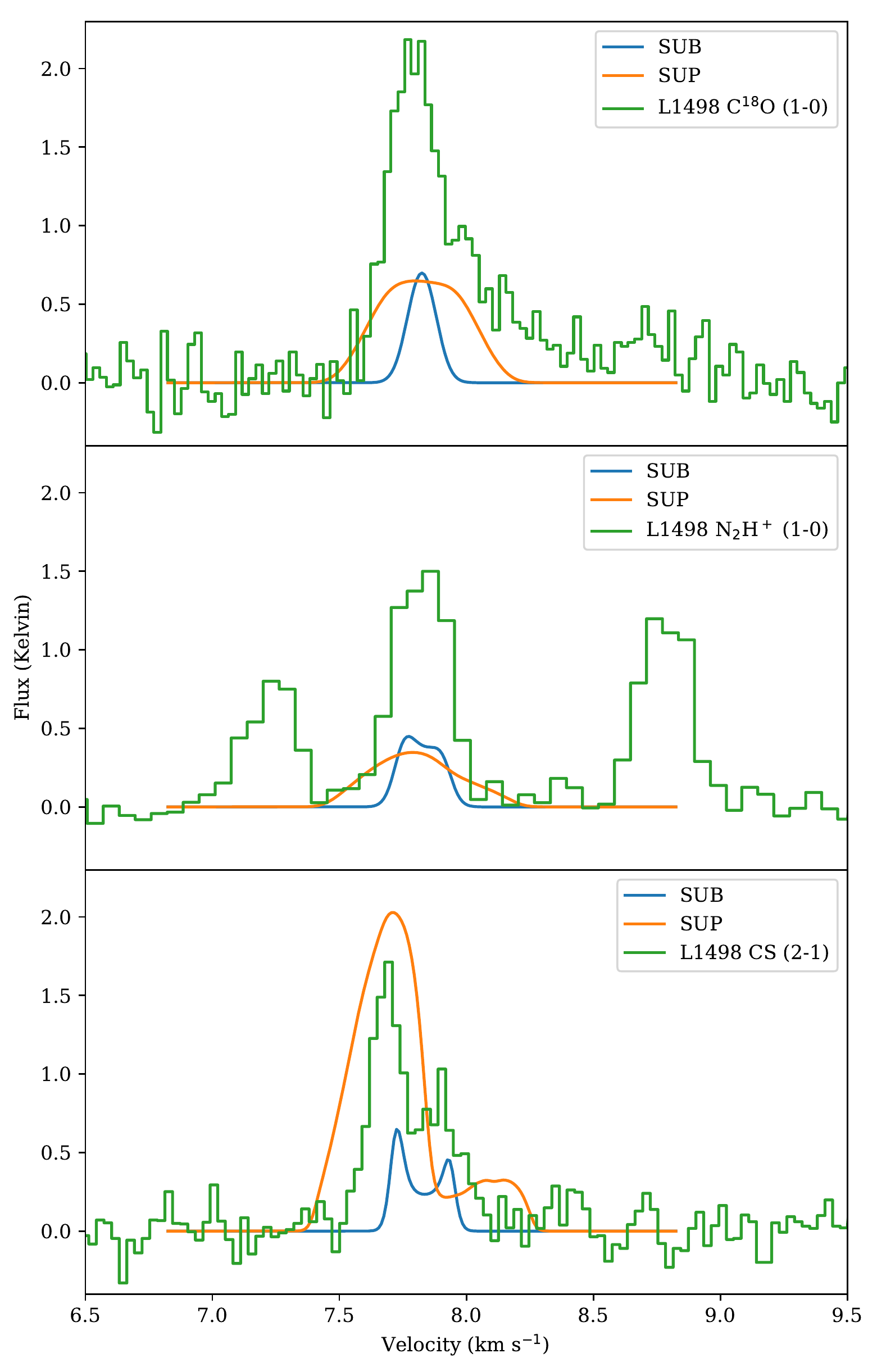}}\quad
	\subfigure{\includegraphics[width=0.45\textwidth]{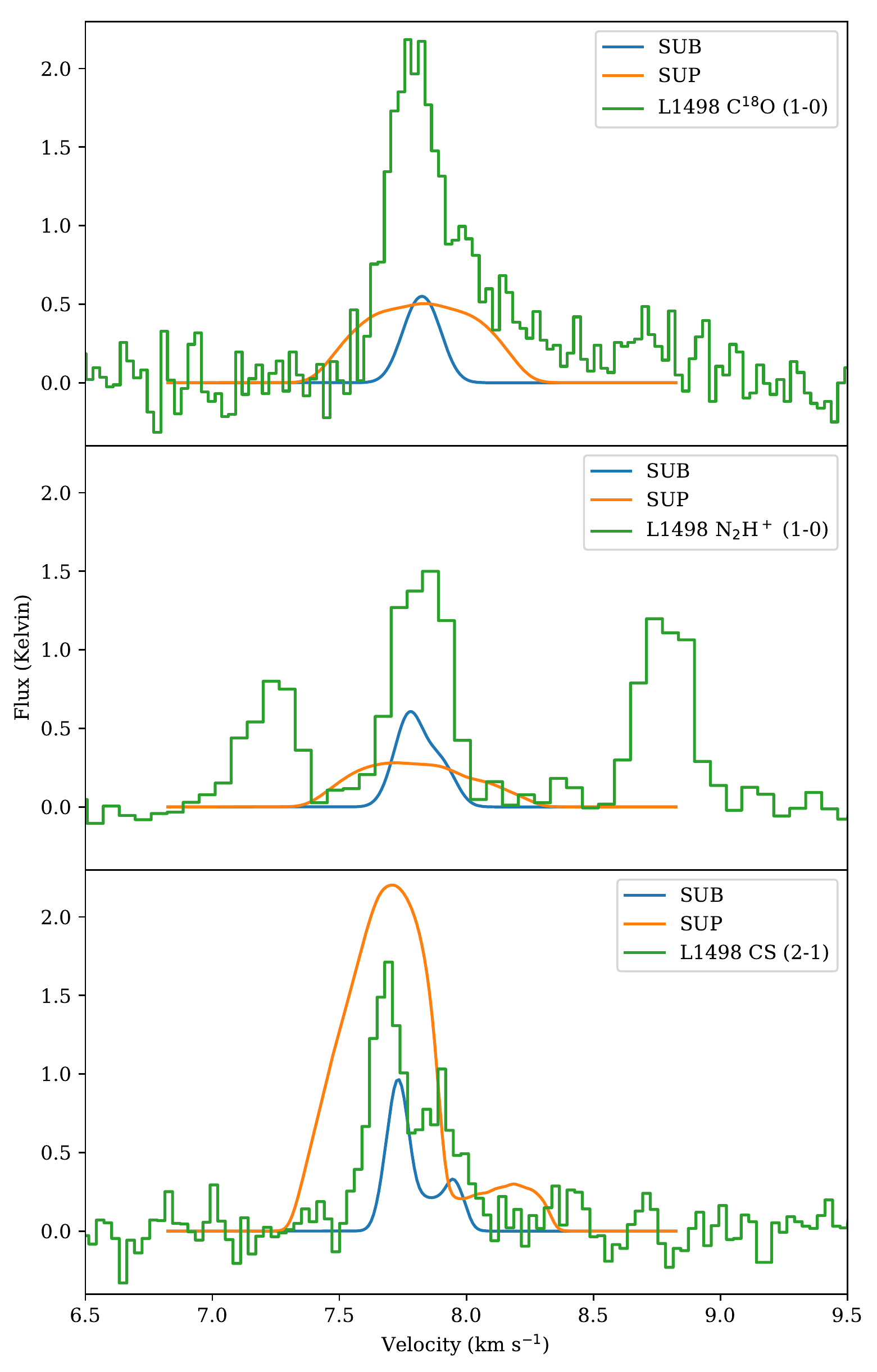}}
	\caption{Comparison of the synthetic line profiles and observations of L1498 presented in \citet{tafalla2002}, assuming a {\bf face-on (left) or side on (right)}.}
	\label{fig:20021498}
\end{figure*}

\begin{figure*}
	\subfigure{\includegraphics[width=0.45\textwidth]{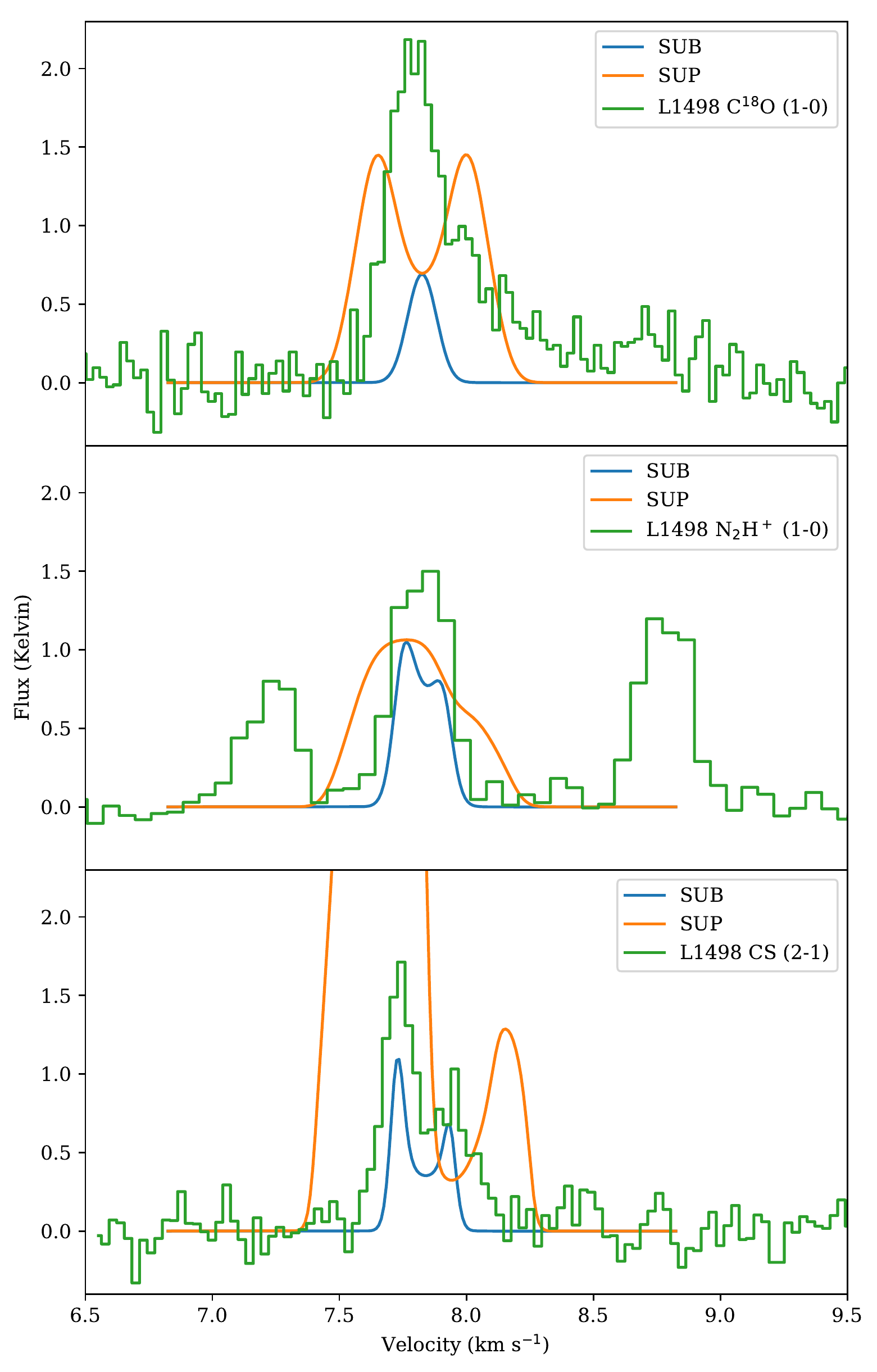}}\quad
	\subfigure{\includegraphics[width=0.45\textwidth]{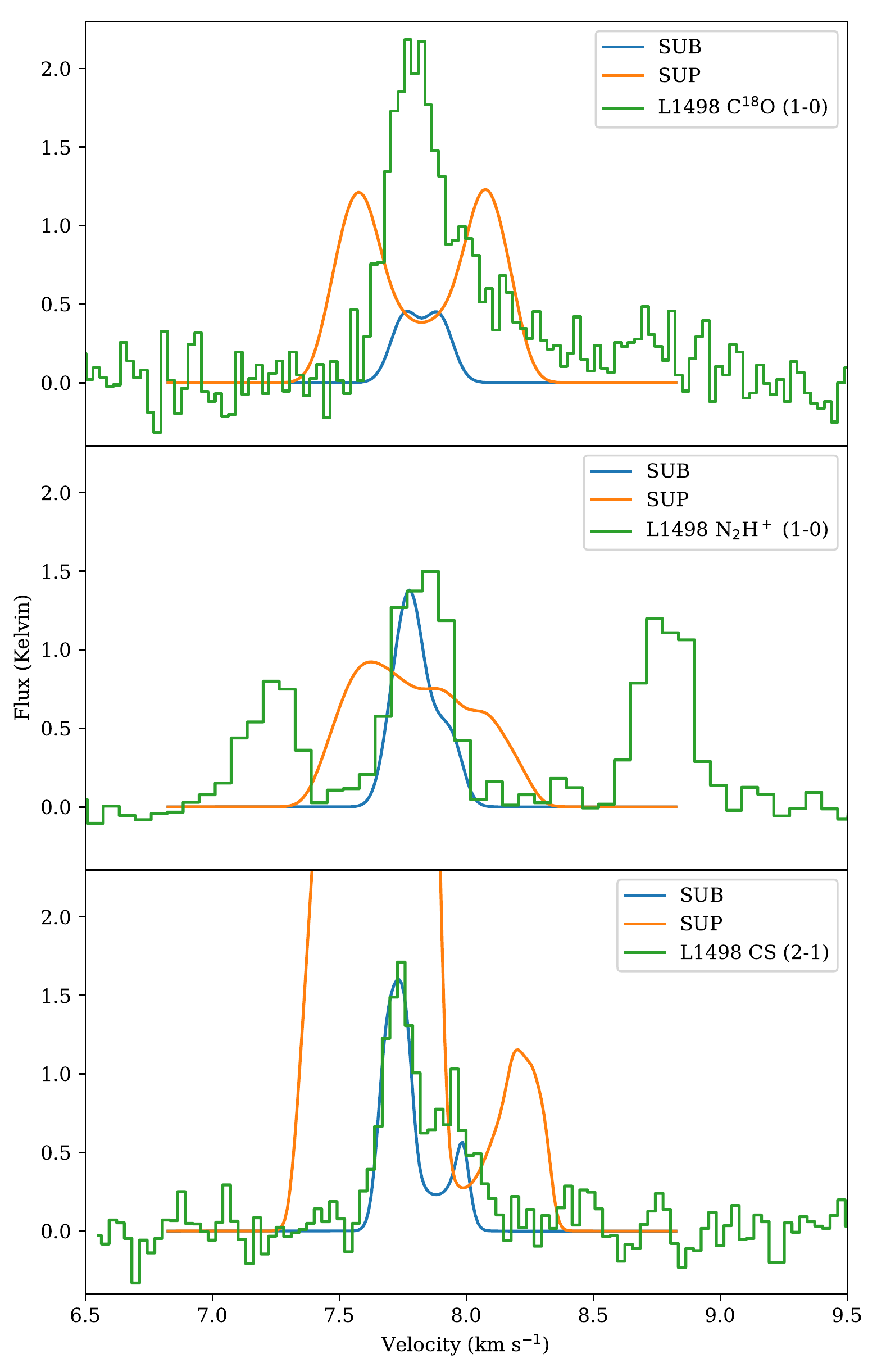}}
	\caption{Comparison of the synthetic line profiles and observations of L1498 presented in \citet{tafalla2002}, assuming a {\bf face-on (left) or side on (right)}. Model line profiles are extracted from the central $0.06 \pc$.}
	\label{fig:20021498_centre}
\end{figure*}

The synthetic lines presented here are from simulations of static, uniform density spheres, and so are not necessarily representative of real prestellar cores. Even restricting ourselves to this model setup, changing the cloud mass or density may alter the line profiles by larger amounts than the differences between sub- and super-critical models. For this reason, we have focused on diagnostics which should not depend on the absolute line intensity, are likely to be least sensitive to initial cloud parameters, and which show a clear dichotomy between the two models. The blue/red peak ratio of the \textbf{CS and HCN} lines, and the intensity ratios \textbf{of those molecules with N$_2$H$^+$}, are easily interpreted as they are \textbf{primarily due} to the longer duration of freeze-out in initially subcritical cores, resulting in less self-absorption and lower peak intensity for CS and HCN while N$_2$H$^+$ is relatively unaffected. Other effects not considered here, such as the cosmic ray ionization rate \citep{wurster2018a} and the initial magnetic field strength \citep{bate2014}, can also change the collapse timescale. However, these effects do not change the collapse timescale to the extent of the difference between super- and sub-critical models \citep{machida2018}. As lower initial densities result in more strongly delayed collapse (the ratio of ionised to neutral species, and thus the ambipolar diffusion timescales, are larger at lower density), the difference in intensity ratio between models discussed here should be even more extreme in this case, while higher initial densities than the $10^4 \pcc$ we assume here seem to be in tension with observed molecular abundances \citep{priestley2019}. It would, however, be useful to confirm whether these signatures are robust to the model parameters, and we intend to conduct a more detailed study of the parameter space in the near future.

With the above caveats, we can compare our model output to observational data as a test of consistency. Of the five prestellar cores studied in \citet{tafalla2002}, only L1498 clearly shows signs of infall in its line profiles, with \citet{kirk2006a} finding the magnetic flux to be slightly super-critical. Therefore we choose L1498 as a test case. The synthetic line profiles for C$^{18}$O, $\rm N_2H^+$, and CS are overlaid on the observed lines for L1498 in Figure \ref{fig:20021498} - we shift the synthetic profile in velocity so that the peak intensity is at the same location as the observed maximum, but otherwise do not modify our data. The observed C$^{18}$O and N$_2$H$^+$ intensities are stronger by a factor of $\sim 2$ than our model line profiles. For CS, the LOW-SUP model has a comparable peak intensity, but a much greater blue/red intensity ratio than observed, whereas the LOW-SUB model is still somewhat weaker than the \citet{tafalla2002} data but with a similar blue/red ratio. The observed N$_2$H$^+$/CS intensity ratio of $\sim 1$ is also in much better agreement with the LOW-SUB model.

L1498 has an observed size of $\sim 100$ arcsec, corresponding to a physical radius of $\sim 0.06 \pc$ for a distance of $140 \pc$ as typically assumed \citep{tafalla2002}. If we instead compare the data with model line profiles from the inner $0.06 \pc$ region of our models, shown in Figure \ref{fig:20021498_centre}, we find the LOW-SUB model provides a good match to the N$_2$H$^+$ and CS profiles regardless of orientation, whereas the LOW-SUP CS intensity is far greater than that observed. We thus favour a subcritical interpretation for the L1498 data. This is not necessarily in conflict with the supercritical mass-to-flux ratio found by \citet{kirk2006a}; their measurement corresponds to the {\it present} value of the mass-to-flux ratio in the dense structure identified as a core, whereas the molecular line data is sensitive to the {\it initial} value in the material which eventually formed L1498.

The LOW-SUB model is a factor $\sim 2$ weaker than the observed C$^{18}$O emission even when only the central region is considered. \citet{tafalla2002} note that there may be contributions to this line from unrelated ambient gas, and the central column density of L1498 ($\sim 10^{22} \pcs$; \citealt{kirk2006a}) is lower than the LOW-SUB model at the point where we produce line profiles ($7.8 \times 10^{22} \pcs$), which may suggest that a more appropriate comparison would be with an earlier, less-depleted phase with correspondingly stronger CO emission. Alternatively, our chosen model parameters may be entirely inappropriate for L1498, and the good agreement between the modelled and observed N$_2$H$^+$ and CS lines is simply a coincidence. Nonetheless, it is clear from Figure \ref{fig:20021498_centre} that the differences between sub- and super-critical line profiles are large enough to be observationally significant, making this method a promising (and complementary) approach to determining the role of magnetic fields in star formation.

\section{Conclusions}

We have post-processed non-ideal MHD simulations of collapsing magnetically sub/supercritical prestellar cores with both a time-dependent chemical network and a line radiative transfer model. The resulting synthetic observations can be directly compared to real data to discriminate between theoretical models of star formation. We find the most promising tracers of collapse mechanism are the ratio between blue and red peaks in the CS $J=2-1$ line profile, and the intensity ratios of lines from molecules affected differently by freeze-out onto grains, such as N$_2$H$^+$ and CS/HCN. Specifically, CS blue/red ratios of $<3$ ($>5$), N$_2$H$^+$/CS ratios of $>0.6$ ($<0.2$), and HCN/N$_2$H$^+$ ratios of $<1$ ($>1$), are indicative of subcritical (supercritical) collapse. These criteria, and the good match with model line profile shapes, suggest that L1498 resulted from subcritical initial conditions, despite being marginally supercritical at present.

\section*{Acknowledgements}

FDP is funded by the Science and Technology Facilities Council. Some figures in this paper were made using SPLASH \citep{price2007a}.

\section*{Data Availability}

The data underlying this article will be shared on reasonable request to the corresponding author.




\bibliographystyle{mnras}
\bibliography{lines}

\begin{thebibliography}{}
\makeatletter
\relax
\def\mn@urlcharsother{\let\do\@makeother \do\$\do\&\do\#\do\^\do\_\do\%\do\~}
\def\mn@doi{\begingroup\mn@urlcharsother \@ifnextchar [ {\mn@doi@}
  {\mn@doi@[]}}
\def\mn@doi@[#1]#2{\def\@tempa{#1}\ifx\@tempa\@empty \href
  {http://dx.doi.org/#2} {doi:#2}\else \href {http://dx.doi.org/#2} {#1}\fi
  \endgroup}
\def\mn@eprint#1#2{\mn@eprint@#1:#2::\@nil}
\def\mn@eprint@arXiv#1{\href {http://arxiv.org/abs/#1} {{\tt arXiv:#1}}}
\def\mn@eprint@dblp#1{\href {http://dblp.uni-trier.de/rec/bibtex/#1.xml}
  {dblp:#1}}
\def\mn@eprint@#1:#2:#3:#4\@nil{\def\@tempa {#1}\def\@tempb {#2}\def\@tempc
  {#3}\ifx \@tempc \@empty \let \@tempc \@tempb \let \@tempb \@tempa \fi \ifx
  \@tempb \@empty \def\@tempb {arXiv}\fi \@ifundefined
  {mn@eprint@\@tempb}{\@tempb:\@tempc}{\expandafter \expandafter \csname
  mn@eprint@\@tempb\endcsname \expandafter{\@tempc}}}

\bibitem[\protect\citeauthoryear{Banerji, Viti, Williams  \& Rawlings}{Banerji
  et~al.}{2009}]{banerji2009}
Banerji M.,  Viti S.,  Williams D.~A.,   Rawlings J. M.~C.,  2009, ]
  {10.1088/0004-637X/692/1/283}, \href
  {https://ui.adsabs.harvard.edu/abs/2009ApJ...692..283B} {692, 283}

\bibitem[\protect\citeauthoryear{Bate, Tricco  \& Price}{Bate
  et~al.}{2014}]{bate2014}
Bate M.~R.,  Tricco T.~S.,   Price D.~J.,  2014, \mn@doi [Monthly Notices of
  the Royal Astronomical Society] {10.1093/mnras/stt1865}, 437, 77

\bibitem[\protect\citeauthoryear{Beltr{\'a}n et~al.,}{Beltr{\'a}n
  et~al.}{2019}]{beltran2019}
Beltr{\'a}n M.~T.,  et~al., 2019, ] {10.1051/0004-6361/201935701}, \href
  {https://ui.adsabs.harvard.edu/abs/2019A\&A...630A..54B} {630, A54}

\bibitem[\protect\citeauthoryear{Brinch \& Hogerheijde}{Brinch \&
  Hogerheijde}{2010}]{brinch2010}
Brinch C.,  Hogerheijde M.~R.,  2010, ] {10.1051/0004-6361/201015333}, \href
  {https://ui.adsabs.harvard.edu/abs/2010A\&A...523A..25B} {523, A25}

\bibitem[\protect\citeauthoryear{Coutens, Commer{\c c}on  \& Wakelam}{Coutens
  et~al.}{2020}]{coutens2020}
Coutens A.,  Commer{\c c}on B.,   Wakelam V.,  2020, ]
  {10.1051/0004-6361/202038437}, \href
  {https://ui.adsabs.harvard.edu/abs/2020A\&A...643A.108C} {643, A108}

\bibitem[\protect\citeauthoryear{Crutcher}{Crutcher}{2012}]{crutcher2012}
Crutcher R.~M.,  2012, ] {10.1146/annurev-astro-081811-125514}, \href
  {https://ui.adsabs.harvard.edu/abs/2012ARA\&A..50...29C} {50, 29}

\bibitem[\protect\citeauthoryear{Crutcher, Hakobian  \& Troland}{Crutcher
  et~al.}{2009}]{crutcher2009}
Crutcher R.~M.,  Hakobian N.,   Troland T.~H.,  2009, ]
  {10.1088/0004-637X/692/1/844}, \href
  {https://ui.adsabs.harvard.edu/abs/2009ApJ...692..844C} {692, 844}

\bibitem[\protect\citeauthoryear{Fiedler \& Mouschovias}{Fiedler \&
  Mouschovias}{1993}]{fiedler1993}
Fiedler R.~A.,  Mouschovias T.~C.,  1993, ] {10.1086/173193}, \href
  {http://adsabs.harvard.edu/abs/1993ApJ...415..680F} {415, 680}

\bibitem[\protect\citeauthoryear{Holdship, Viti, {Jim{\'e}nez-Serra},
  Makrymallis  \& Priestley}{Holdship et~al.}{2017}]{holdship2017}
Holdship J.,  Viti S.,  {Jim{\'e}nez-Serra} I.,  Makrymallis A.,   Priestley
  F.,  2017, ] {10.3847/1538-3881/aa773f}, \href
  {https://ui.adsabs.harvard.edu/abs/2017AJ....154...38H} {154, 38}

\bibitem[\protect\citeauthoryear{Jiang, Li  \& Fan}{Jiang
  et~al.}{2020}]{jiang2020}
Jiang H.,  Li H.-b.,   Fan X.,  2020, \mn@doi [The Astrophysical Journal]
  {10.3847/1538-4357/ab672b}, 890, 153

\bibitem[\protect\citeauthoryear{Kirk, {Ward-Thompson}  \& Crutcher}{Kirk
  et~al.}{2006}]{kirk2006a}
Kirk J.~M.,  {Ward-Thompson} D.,   Crutcher R.~M.,  2006, \mn@doi [Monthly
  Notices of the Royal Astronomical Society]
  {10.1111/j.1365-2966.2006.10392.x}, 369, 1445

\bibitem[\protect\citeauthoryear{Lee, Roueff, {Pineau des Forets}, Shalabiea,
  Terzieva  \& Herbst}{Lee et~al.}{1998}]{lee1998}
Lee H.-H.,  Roueff E.,  {Pineau des Forets} G.,  Shalabiea O.~M.,  Terzieva R.,
    Herbst E.,  1998, \href
  {http://adsabs.harvard.edu/abs/1998A\%26A...334.1047L} {334, 1047}

\bibitem[\protect\citeauthoryear{Lin, Pagani, Lai, Lef{\`e}vre  \& Lique}{Lin
  et~al.}{2020}]{lin2020}
Lin S.-J.,  Pagani L.,  Lai S.-P.,  Lef{\`e}vre C.,   Lique F.,  2020, ]
  {10.1051/0004-6361/201936877}, \href
  {https://ui.adsabs.harvard.edu/abs/2020A\&A...635A.188L} {635, A188}

\bibitem[\protect\citeauthoryear{Lippok et~al.,}{Lippok
  et~al.}{2013}]{lippok2013}
Lippok N.,  et~al., 2013, ] {10.1051/0004-6361/201322129}, \href
  {https://ui.adsabs.harvard.edu/abs/2013A\&A...560A..41L} {560, A41}

\bibitem[\protect\citeauthoryear{Machida, Higuchi  \& Okuzumi}{Machida
  et~al.}{2018}]{machida2018}
Machida M.~N.,  Higuchi K.,   Okuzumi S.,  2018, \mn@doi [Monthly Notices of
  the Royal Astronomical Society] {10.1093/mnras/stx2589}, 473, 3080

\bibitem[\protect\citeauthoryear{McElroy, Walsh, Markwick, Cordiner, Smith  \&
  Millar}{McElroy et~al.}{2013}]{mcelroy2013}
McElroy D.,  Walsh C.,  Markwick A.~J.,  Cordiner M.~A.,  Smith K.,   Millar
  T.~J.,  2013, ] {10.1051/0004-6361/201220465}, \href
  {http://adsabs.harvard.edu/abs/2013A\%26A...550A..36M} {550, A36}

\bibitem[\protect\citeauthoryear{Mouschovias}{Mouschovias}{1976}]{mouschovias1976}
Mouschovias T.~C.,  1976, \mn@doi [The Astrophysical Journal] {10.1086/154436},
  206, 753

\bibitem[\protect\citeauthoryear{Oberg \& Bergin}{Oberg \&
  Bergin}{2020}]{oberg2020}
Oberg K.~I.,  Bergin E.~A.,  2020, arXiv e-prints, \href
  {https://ui.adsabs.harvard.edu/abs/2020arXiv201003529O} {p. arXiv:2010.03529}

\bibitem[\protect\citeauthoryear{Pagani, Lesaffre, Jorfi, Honvault,
  {Gonz{\'a}lez-Lezana}  \& Faure}{Pagani et~al.}{2013}]{pagani2013}
Pagani L.,  Lesaffre P.,  Jorfi M.,  Honvault P.,  {Gonz{\'a}lez-Lezana} T.,
  Faure A.,  2013, ] {10.1051/0004-6361/201117161}, \href
  {http://adsabs.harvard.edu/abs/2013A\%26A...551A..38P} {551, A38}

\bibitem[\protect\citeauthoryear{Price}{Price}{2007}]{price2007a}
Price D.~J.,  2007, \mn@doi [Publications of the Astronomical Society of
  Australia] {10.1071/AS07022}, 24, 159

\bibitem[\protect\citeauthoryear{Price et~al.,}{Price et~al.}{2018}]{price2018}
Price D.~J.,  et~al., 2018, ] {10.1017/pasa.2018.25}, \href
  {http://adsabs.harvard.edu/abs/2018PASA...35...31P} {35, e031}

\bibitem[\protect\citeauthoryear{Priestley, Viti  \& Williams}{Priestley
  et~al.}{2018}]{priestley2018}
Priestley F.~D.,  Viti S.,   Williams D.~A.,  2018, ]
  {10.3847/1538-3881/aac957}, \href
  {http://adsabs.harvard.edu/abs/2018AJ....156...51P} {156, 51}

\bibitem[\protect\citeauthoryear{Priestley, Wurster  \& Viti}{Priestley
  et~al.}{2019}]{priestley2019}
Priestley F.~D.,  Wurster J.,   Viti S.,  2019, ] {10.1093/mnras/stz1869},
  \href {https://ui.adsabs.harvard.edu/abs/2019MNRAS.488.2357P} {488, 2357}

\bibitem[\protect\citeauthoryear{Sch{\"o}ier, {van der Tak}, {van Dishoeck}  \&
  Black}{Sch{\"o}ier et~al.}{2005}]{schoier2005}
Sch{\"o}ier F.~L.,  {van der Tak} F. F.~S.,  {van Dishoeck} E.~F.,   Black
  J.~H.,  2005, ] {10.1051/0004-6361:20041729}, \href
  {http://adsabs.harvard.edu/abs/2005A\%26A...432..369S} {432, 369}

\bibitem[\protect\citeauthoryear{Soam et~al.,}{Soam et~al.}{2018}]{soam2018}
Soam A.,  et~al., 2018, ] {10.3847/1538-4357/aac4a6}, \href
  {https://ui.adsabs.harvard.edu/abs/2018ApJ...861...65S} {861, 65}

\bibitem[\protect\citeauthoryear{Soam et~al.,}{Soam et~al.}{2019}]{soam2019}
Soam A.,  et~al., 2019, ] {10.3847/1538-4357/ab39dd}, \href
  {https://ui.adsabs.harvard.edu/abs/2019ApJ...883...95S} {883, 95}

\bibitem[\protect\citeauthoryear{Tafalla, Myers, Caselli, Walmsley  \&
  Comito}{Tafalla et~al.}{2002}]{tafalla2002}
Tafalla M.,  Myers P.~C.,  Caselli P.,  Walmsley C.~M.,   Comito C.,  2002, ]
  {10.1086/339321}, \href {http://adsabs.harvard.edu/abs/2002ApJ...569..815T}
  {569, 815}

\bibitem[\protect\citeauthoryear{Tafalla, {Santiago-Garc{\'i}a}, Myers,
  Caselli, Walmsley  \& Crapsi}{Tafalla et~al.}{2006}]{tafalla2006}
Tafalla M.,  {Santiago-Garc{\'i}a} J.,  Myers P.~C.,  Caselli P.,  Walmsley
  C.~M.,   Crapsi A.,  2006, ] {10.1051/0004-6361:20065311}, \href
  {http://adsabs.harvard.edu/abs/2006A\%26A...455..577T} {455, 577}

\bibitem[\protect\citeauthoryear{Tassis \& Mouschovias}{Tassis \&
  Mouschovias}{2004}]{tassis2004a}
Tassis K.,  Mouschovias T.~C.,  2004, \mn@doi [The Astrophysical Journal]
  {10.1086/424901}, 616, 283

\bibitem[\protect\citeauthoryear{Tassis, Willacy, Yorke  \& Turner}{Tassis
  et~al.}{2012a}]{tassis2012}
Tassis K.,  Willacy K.,  Yorke H.~W.,   Turner N.~J.,  2012a, ]
  {10.1088/0004-637X/753/1/29}, \href
  {http://adsabs.harvard.edu/abs/2012ApJ...753...29T} {753, 29}

\bibitem[\protect\citeauthoryear{Tassis, Willacy, Yorke  \& Turner}{Tassis
  et~al.}{2012b}]{tassis2012b}
Tassis K.,  Willacy K.,  Yorke H.~W.,   Turner N.~J.,  2012b, ]
  {10.1088/0004-637X/754/1/6}, \href
  {http://adsabs.harvard.edu/abs/2012ApJ...754....6T} {754, 6}

\bibitem[\protect\citeauthoryear{Tassis, Willacy, Yorke  \& Turner}{Tassis
  et~al.}{2014}]{tassis2014}
Tassis K.,  Willacy K.,  Yorke H.~W.,   Turner N.~J.,  2014, \mn@doi [Monthly
  Notices of the Royal Astronomical Society] {10.1093/mnrasl/slu130}, 445, L56

\bibitem[\protect\citeauthoryear{Tritsis, Panopoulou, Mouschovias, Tassis  \&
  Pavlidou}{Tritsis et~al.}{2015}]{tritsis2015}
Tritsis A.,  Panopoulou G.~V.,  Mouschovias T.~C.,  Tassis K.,   Pavlidou V.,
  2015, \mn@doi [Monthly Notices of the Royal Astronomical Society]
  {10.1093/mnras/stv1133}, 451, 4384

\bibitem[\protect\citeauthoryear{Wurster}{Wurster}{2016}]{wurster2016}
Wurster J.,  2016, ] {10.1017/pasa.2016.34}, \href
  {http://adsabs.harvard.edu/abs/2016PASA...33...41W} {33, e041}

\bibitem[\protect\citeauthoryear{Wurster, Bate  \& Price}{Wurster
  et~al.}{2018}]{wurster2018a}
Wurster J.,  Bate M.~R.,   Price D.~J.,  2018, ] {10.1093/mnras/stx3339}, \href
  {http://adsabs.harvard.edu/abs/2018MNRAS.475.1859W} {475, 1859}

\makeatother
\end{thebibliography}
\vspace{10pt}
This article has been accepted for publication in the Monthly Notices of the Royal Astronomical Society published by Oxford University Press on behalf of the Royal Astronomical Society.




\bsp	
\label{lastpage}
\end{document}